# Degradation-Reducing Control for Dynamically Reconfigurable Batteries

Tomas Kacetl, Jan Kacetl, Nima Tashakor, Jingyang Fang, Malte Jaensch, and Stefan Goetz

*Abstract*—**Cascaded circuits such as modular multilevel converters (MMC) offer attractive qualities in reconfigurable battery applications. In contrast to conventional hard-wired dc battery packs, the MMC topology loads modules with ac current, which may lead to additional ageing of batteries. As recent studies reveal, such ageing of batteries occurs at low-frequency load ripple, and almost vanishes at high frequencies. State of the art in MMC battery control focuses on state of charge and temperature balancing of individual modules. Previous methods to suppress ripple rely on slow feedback loops and low dynamics, which tends to form low-frequency patterns in the module load that negatively contribute to their ageing. This paper presents a novel module-current-oriented high-bandwidth control technique which minimizes low-frequency components in the module load spectrum. The control method respects limitations related to module data acquisition and enhances the feedback bandwidth using observation techniques. We verify the proposed method experimentally on a laboratory setup and estimate the influence on the battery cells.**

## I. INTRODUCTION

ELECTROMOBILITY is a dynamically developing field of power electronics and battery applications, and so is grid storage. Both use battery packs as an energy tank and a semiconductor inverter to generate the ac output for the motor or grid. The conventional battery of an electric vehicle wires cells in a battery pack with certain fixed parallel and serial configuration, where the number of series cells determines the terminal voltage and the number of parallel cells the current capability.

One of the issues of hard-wired batteries is balancing of individual cells [1]. Tolerances and the corresponding spread in battery cell parameters cause differences in discharging of individual cells of the battery pack. The whole battery pack then follows Liebig's law of the minimum (barrel theory) with respect to each of these parameters, where the overall capacitance of the battery is limited by the cell of minimal capacitance. Similarly the weakest cell with respect to max power, capacity loss, and individual defects and associated differential degradation limits the entire pack. In addition, hard wiring of the cells in the battery pack offers minimal fault tolerance. The failure of a single cell typically results in failure of the entire battery pack and the electric vehicle subsequently. Further, the urge to increase system efficiency and reduce charging time of electric vehicles has resulted in a steady increase of the battery pack voltage. However, increasing battery voltages requires corresponding blocking capability of semiconductors in inverters, which limits scalability of this approach.

Alternatively, modular circuit structures such as the modular multilevel converters (MMC) can integrate batteries [2]. The MMC-battery topology combines power electronics and batteries, typically based on cascaded H bridges (CHB). Although solutions breaking the battery pack down to the cell level have been suggested, the advantages can be exploited already at coarser partitioning so that each module contains several hard-wired cells. The multilevel approach allows the use of low-voltage semiconductors in the system and offers scalability [3-5]. In contrast to hard-wired batteries, the distributed power electronics can control the power of individual modules [6]. Thus, reconfigurable battery systems offer excellent balancing of the state of charge [7, 8] and state of health [9-12], introduce fault tolerance by bypassing defective modules or even semiconductors [13-16], and increase the effectively available capacitance of battery systems [7, 17]. In addition, reconfigurable systems exhibit high efficiency, low cost, and modularity [4, 18, 19]. However, in comparison to hard-wired dc batteries, MMC-based topologies with ac load expose the battery cells to load ripple of twice the ac output frequency and of switching [20-23].

In addition to grid storage requiring enhanced balancing performance for poorly matched second-life batteries [6, 13, 24, 25], electric vehicles benefit from the advantages of reconfigurable battery systems [26-28]. Among different MMC topologies, cascade double H bridges (CHB$^2$) offer series and parallel connection modes between modules, also often denoted as MMSPC [6, 14, 29-31]. CHB[2] promise better load distribution among battery modules, higher failure safety, lower effective source impedance, and lower ripple load are major advantages particularly for battery applications [21, 22, 32, 33].

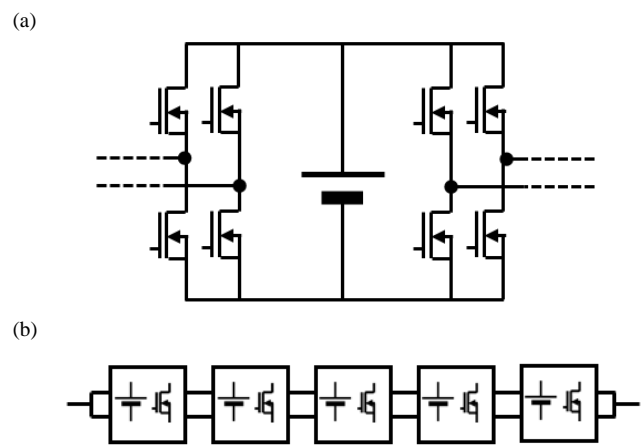

**Figure 1.** Diagram of the system topology: (a) battery module with HB$^2$ switch topology allowing paralleling of modules, (b) phase string topology of the MMSPC converter with integrated batteries.

Section II explains the sources of ripple current and its impact on batteries. Section III discusses known control strategies. Section IV introduces our novel control strategy for mitigating factors leading to accelerated ageing of the battery cells, which Section VI underpins with experimental results.

## II. INFLUENCE OF RIPPLE-CURRENT ON BATTERIES

### A. Module current

For both electric vehicles and grid-connected storage with hard-wired batteries, the dominant load is determined by the inverter. The ac side of the inverter supplies the grid or an electric motor with ac current, whereas the dc link of the inverter loads the battery with a current resulting from the operation of the inverter [34]. For three-phase output, the dc-link load typically contains a dc component with the $6^{th}$ harmonic as well as further multiples of the ac frequency [21, 35, 36]. In addition, the spectrum includes the switching frequency of the inverter and its harmonics. The distribution of the current in the battery pack among individual cells or sub-batteries is determined by the impedance of interconnections and cells, and their voltage differences. Different impedances of current paths through the battery can lead to unequal discharge rates and consequently to a spread in cell ageing [37-40].

In MMSPC, module currents depend on the macro-level topology. In case of a star connection, the modules are divided into phase strings, where each of the phase strings supplies one output phase. As a result, the spectrum of the module load contains a strong $2^{nd}$ harmonic of the output ac frequency. Further, MMCs without parallel module connectivity alternate between an active state of the module (series module connection), where modules take phase load, and an inactive state of the module (bypass state), where modules take zero load. Alternating these states introduces components of variable frequency in the module load spectrum, which depend on the specific control strategy. In the MMSPC topology, the parallel states may substitute the inactive bypass state [41]. Paralleling modules eliminates no-load states and lowers the load of the active modules to bring both closer to the mean. Consequently, utilization of the parallel state increases the dc component of the load spectrum, and further expands the possibilities of a module load control.

Load ripple is commonly subject to hardware filtering. The conventional topology of electric vehicles uses dc-link capacitors to filter out switching frequencies and the load ripple. Similarly, filtering often serves for the reduction of the load ripple in MMC modules [42]. Furthermore, a voltage controller can systematically inject harmonic voltage in multi-phase systems to even out currents across the individual phase and consequently reduce the module load ripple [21, 43].

In addition to these solutions, MMSPC topologies allow a reduction of the module load ripple through the parallel connection of neighboring modules [32]. The phase-string topology with a shared double neutral point further allows module paralleling across this double neutral point to exchange power and improve the load distribution [22, 41, 44]. The load ripple reduction can be an objective of the control algorithm, as will be presented further later on.

### B. Lithium cell degradation

Lithium-ion cells of the various sub-types offer high power and energy density, which renders the technology suitable for portable devices and electric vehicles. Further performance improvement of the technology is a subject of research in electrochemistry and material engineering. During operation, the elements and materials of the battery cells undergo degradation processes so that the cell gradually loses its capacity and performance. The degradation is a result of many physical and electrochemical processes, also called faradaic processes, connected to charging and discharging routines of the cell and the operating conditions. Industry pays attention to high discharge and charge rates. Among the most relevant degradation modes connected to high current or pulse load are solid electrolyte interface (SEI) growth and SEI decomposition, graphite exfoliation, structural disordering, loss of electrical contact between electrodes and terminals, particle cracking and island formation [45-50]. Each of these processes results in either loss of cycleable lithium or loss of active electrode material.

Nevertheless, charging and discharging of a battery cell does not necessarily lead to faradaic processes. Electrically charged electrodes and adjacent electrolyte form a charge double layer [51-54]. Due to the small spacing of relatively large charges, the capacitance of the double layer can provide enough charge for short current pulses without further chemical reactions.

Electrochemical reactions have limited kinetics, often determined by diffusion, and electrically appear as low-pass system. To initiate the chemical reactions, a high current has to be maintained for an extended period of time in the range of milliseconds [55]. Electrochemical models show, that short pulses are almost completely buffered with double-layer capacitance at the electrode–electrolyte interface and the electrolyte capacitance, which has a large dielectric constant as a byproduct of large polarity for increased lithium solubility [56]. The impedance of electrochemical processes and battery cells is further studied in [57, 58].

The actual contribution of the rippled load to accelerated ageing is likewise receiving more attention in recent experiments. Bessman et al. contradict concerns over accelerated ageing, and show actual influence to be minimal, especially for higher frequencies [59]. Results reported in previous experiments mainly support their conclusion and assume existence of a corner frequency or transition band, above which the dielectric charge absorption capability of electrodes leads to sudden decrease in battery ageing [60]. Thus, minimizing low-frequency content of a battery load spectrum is the main factor for proper treatment of battery cells to minimize accelerated ageing, which also complies with the most recent experiment particularly focusing on MMC load [61]. This observation is the foundation of our control approach.

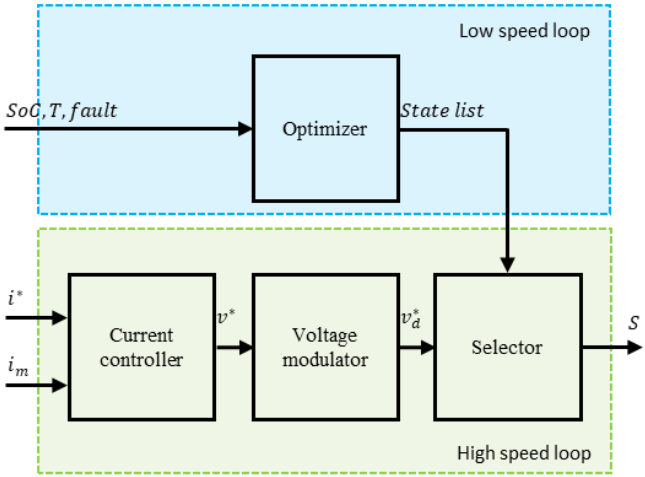

**Figure 2.** Parallel asynchronous optimization for MMC control, where a low-speed loop optimizes module states or state transitions and stores them in a list or look-up table, which a high-speed loop of the actual controller, modulator, and scheduler uses to actuate the transistors in the modules.

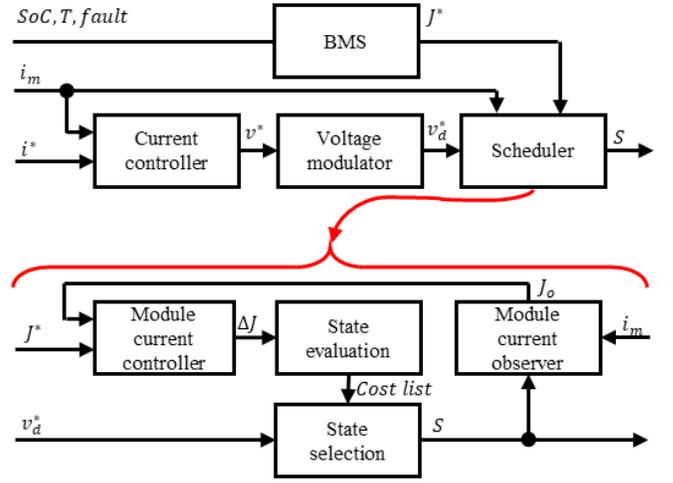

**Figure 3.** Block diagram of proposed MMC control algorithm with fast quasi-inline optimization inside the scheduler block, which is further detailed at the bottom.

## III. STATE OF THE ART IN CONTROL

MMC topologies, including MMSPC, incorporate low-voltage semiconductor switches into each module. The higher number of individually governable active components challenges the control. In contrast to hard-wired batteries, however, the MMC provides control over individual battery modules and their power flow. In addition to a direction of the current, topologies with a parallel mode further control distribution of the phase power among all available modules. Primarily, the control algorithm provides a demanded output voltage. The output voltage is controlled by switching a proper number of phase modules to the series state. The secondary objective of the control algorithm is a load of individual modules. The load distribution is typically controlled with respect to a charge (resp. voltage) imbalance among string modules, and might be further used to follow secondary objectives. [62-69].

In addition to series and bypass states of MMC and CHB, modules of the MMSPC and CHB² provide also a parallel module interconnection state. The parallel state provides further degrees of freedom for the controller to optimize additional objectives beyond the output voltage/current and voltage balancing. Examples for additional objectives are load ripple reduction and phase impedance control. Phase-shifted carrier control, for instance, can use the parallel mode to maximize utilization of the modules [33, 70, 71].

However, state of charge (SoC) or voltage balancing alone may be sufficient for modules with capacitors, but not with battery cells. Battery cells require control of the individual load current's absolute amplitude and ideally also ripple. A recently presented approach for controlling those expands above optimization approaches of states and includes criteria such as SoC, temperature, phase current, and the previous state for the selection of the next state [72].

Different dynamics of output voltage modulation versus the discharge rates of batteries justifies separation of the optimization into two parts running on a controller at different speeds to reduce computational effort. A fast loop of the controller selects proper module state configurations from an optimized list provided by the slower loop with more time for the computational optimization and no need to face deadlines associated with the fast real-time part (see Fig. 2).

The optimal states are selected with respect to state of charge weighted by estimated load J, where calculated state cost (SC) may optionally further includes ripple reduction term, as

$$SC(s) = \sum_{i=1}^{N} J_i(s) \cdot (SoC_i - SoC_{\mathrm{mean}}) + J_i^2(s).$$

(1)

Whereas the load distribution is partly reflected in the state selection, the actual waveform of the load remains uncontrolled. Specht et al. suggest updating the content of the optimized list of states once every second, which leaves the content constant during the update period and introduces constant switching patterns projected to the module load. Any imbalance of controlled values results in either extensive utilization or mitigation of the module load in the switching pattern. Formation of temporal patterns can be especially harmful, when reaching the millisecond range as outlined about the faradaic dynamics in the introduction of our paper. Increasing the update period is limited by relying on typically slow acquisition of the feedback data and bandwidth of communication buses, which rarely allows sufficiently fast feedback.

## IV. SUGGESTED CONTROL

To achieve optimal battery treatment through the module load, the distribution of the phase current becomes the main objectives of our scheduling algorithm. The algorithm employs a structure depicted in Fig. 3. The structure comprises a set of current controllers (regulators) assigned to each module, an optimization routine for the selection of a proper system configuration, and an observer, which closes the control loop. For illustration, Figure 3 shows the whole phase current control loop with a phase current controller and a voltage modulator. Our control scheduling algorithm does not rely on any of these components; so the following will treat them as given and refers to the literature for details [73].

More importantly, our algorithm aims for a maximization of the bandwidth, which should prevent any switching patterns and contribute to increased life time of the battery cells. Efficient suppression of switching patterns lies in matching the module control loop and the phase control loop in speed. Therefore, in contrast to previous suggestions [72], all blocks of the module controller shall preferably run within a switching period of the phase control loop. Nevertheless, the module control loop is not yet time-critical, since any delay in execution only introduces a small switching pattern with length respective to the delay. Indeed, any loop execution time under the period of the phase load contributes to decrease of the load frequency content.

### A. Module current controller

The module current controller guarantees discharging (resp. charging) modules with the demanded current and provides an interface for any higher-level entity. Whereas proper execution time of the controller is essential for desired functionality of the scheduling approach, the bandwidth of the controller is preferably small. The requirement follows from discrete distribution of the phase load among modules, where higher filtering of the controlled current provides enough time for the optimization routine to switch an appropriate sequence of string states to provide the demanded current on average. A typical implementation of the current controller may be a PI controller.

### B. Module current observer

The control loop requires considerably fast feedback to run at maximum speed. An acquisition of the module current and transmission of the measured value to the controller represents either unacceptable propagation delay or heavy load of a data bus. Typical period of the data acquisition is in order of milliseconds, which is comparable to the load frequency and may not be sufficient for proper control of the module load frequency. Our controller architecture solves the slacking feedback using an observation technique.

The actual current load $i_{bi}$ of module $i$ is a result of the phase string configuration (series–parallel configuration of modules), current load of the phase $I_L$, and voltage $V_B$ as well as impedance ratios of modules $R_B$ vs. their interconnection paths $R_S$. In principle, the string configuration comprises a set of parallel groups, where each parallel group forms one level of the output voltage and is loaded by the phase current. Equation set

$$\begin{bmatrix} R_{D1} & R_{U2} & 0 & 0 & 0 \\ R_{L1} & R_{D2} & R_{U3} & 0 & 0 \\ R_{L2} & R_{L2} & R_{D3} & R_{U4} & 0 \\ R_{L3} & R_{L3} & R_{L3} & R_{D4} & R_{U5} \\ 1 & 1 & 1 & 1 & 1 \end{bmatrix} \cdot \begin{bmatrix} i_{b1} \\ i_{b2} \\ i_{b3} \\ i_{b4} \\ i_{b5} \end{bmatrix} = \begin{bmatrix} B_1 \\ B_2 \\ B_3 \\ B_4 \\ I_L \end{bmatrix}$$
$$R_{Dj} = -(R_{Bj} + R_{SLj} + R_{SH1})$$
$$R_{Lj} = -(R_{SHj} + R_{SLj})$$
$$R_{Uj} = R_{Bj}$$
$$B_j = V_{Bj+1} - V_{Bj} - R_{SLj} \cdot I_L$$
$$\text{(2)}$$

governs further distribution of the phase current among paralleled modules, where the dimension of the problem is equal to the number of parallel modules. Considering constant values of all resistances allows pre-calculation of the impedance matrix and significantly simplifies the algorithm complexity solving (2). The current distribution is then reduced to a function of module voltage differences, which are—through the effort of the controller—kept minimal and develop only slowly over time dependent on the battery capacity. Under the assumption of constant module voltages, the whole problem can be further simplified: a look-up table, where the observer pre-estimates the expected module current distribution of each feasible string configuration in the look-up table. Values in the look-up table may occasionally update once the governing quantities noticeably change.

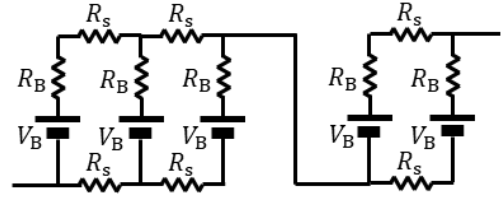

**Figure 4.** Example string configuration of the examined string state $S$ of Eq. (3).

However, even an entirely constant look-up table can temporally approximate the module load, which sufficiently substitutes missing feedback within the acquisition period. In the simplest embodiment of the current observer, the values in the look-up table correspond to the ideal share of the load among paralleled modules. In fact, operating conditions of the system are usually not too far to meet the ideal share per module. First, balance of modules is an objective of the system controller and a properly controlled system guarantees voltage balance. Second, the resistance ratio between electronic switches and battery cells typically stays high enough to keep the error on the order of few percents. In addition, the cumulative error of the ideal current share estimation is further compensated through symmetrical alternation of phase string configurations for the positive and negative half-period of the output voltage. Alternatively, any higher-level entity may ultimately cancel the error by projection of the error onto the controller reference of the module load.

We further demonstrate an example of the observation technique on a string configuration in Figure 4. The figure displays the resultant configuration of the examined converter string state S. The string configuration corresponds to the state of individual modules per

$$S = (P, P, S_+, P, S_+).$$
$$\text{(3)}$$

The state $S$ connects modules such that Modules 1 to 3 form a first and Modules 4 and 5 a second parallel group. The last element of the state vector controls the output terminals of the string. Since none of the groups of state $S$ bypasses the load, the distribution $\vec{J_m}$ can be approximated by the inverse of the number of modules in each parallel group:

$$\vec{J_m}(S) = (0.\overline{3}, 0.\overline{3}, 0.\overline{3}, 0.5, 0.5)$$
$$\text{(4)}$$

The current distribution is multiplied by the magnitude of the actual phase current and provided to the output of the observer block at every occurrence of the state $S$. Similarly, the look-up table of the current observer defines distribution $J_m$ to each feasible string state.

### C. State optimization

The optimization routine deals with bounds of the parallel state. Selection of the optimal state is subject to individual demands of the module current controller. While topologies without parallel state can easily handle the controller demands by sorting algorithms and prioritized active/inactive states for modules with high/low regulator demand, topologies with parallel state need to deal with an increased number of string states. To equally distribute the phase load, the majority of modules preferably stay in active state and rather control their contribution to the phase current by appropriate clustering in parallel groups. Similarly to the current observer, all states in the optimization block are represented by the current distribution. The optimization routine selects one state with optimal current distribution $J_{i,m}$ respecting the demand of the module current controllers $J_i^*$.

Our algorithm uses a typical least-square criterion of optimality to evaluate each state per

$$SC(s) = \sum_{i=1}^{N} \left( J_i^* - J_{i,m}(s) \right)^2. \tag{5}$$

The least square criterion guarantees sufficient effort to meet the regulator demands and simultaneously reduces conduction losses by preventing states with outlying load distribution (e.g., extensive utilization of bypassing).

To achieve a sufficient bandwidth of the optimization routine, we constraint transients between consecutive string configurations by limiting the number of switches that toggle in each step. In other words, to increase the phase voltage any parallel (or bypass) connection of the actual state is switched to series connection, and to decrease the phase voltage any series connection of the actual state is switched to parallel (or bypass).

## V. Demonstration

The proposed method is experimentally evaluated on a laboratory test setup. The experiment compares current load of our method to the method proposed in [72]. The aim of the demonstration is to emphasize importance of the fast feedback and control loop in prevention of switching pattern formation.

The test setup comprises five MMSPC modules with CHB[2]

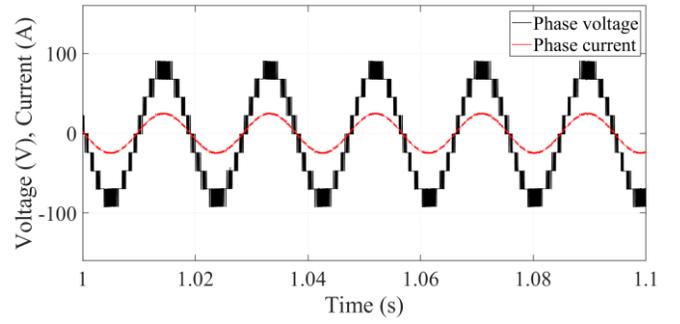

**Figure 6.** Experimental load of the eleven-level MMSPC phase string ($m = 0.7$, $I_{pk} = 25$ A).

| Load parameter | Value |
| --- | --- |
| Phase current | $10 - 35$ A |
| Modulation index | $0.3 - 1.0$ |
| Load frequency | 50 Hz |
| Switching rage | 20 kHz |

topology presented in Figure 1. The module is assembled with two parallel silicon MOSFETs (100 V, 1.5 mΩ, IAUT300-N10S5N015, Infineon) for every switch in the topology, and ceramic capacitors with a total capacitance of 490 µF on the dc bus. The dc bus of each module is supplied by a LiFePO$_4$ battery (22.5 V, 6s, 6.2 Ah).

The system controller uses a Mars ZX3 module assembled with Xilinx's Zynq-7020 system-on-chip, Enclustra). The control algorithm runs fully on the FPGA part. The module power electronics boards offer direct access to gate signals and the controller assigns a GPIO to every switch in the system. The setup implements a sigma-delta modulator running at 20 kHz.

### A. Sensorless operation

To avoid current sensors in each module and fast sampling, we choose to adapt our algorithm for sensorless operation, which sufficiently demonstrates strengths of our scheduling approach. Few modifications in Figure 5 implement such sensorless operation. As a result, the scheduler controls and presumably guarantees equal utilization of battery modules.

First, we change the module current observer to provide utilization feedback, which is achieved by using the signum of the modulated voltage to feed the phase current terminals of the observer block. Considering unit power factor, the signum respects direction of the phase current, and the current distribution $J_{\rm m}$ (see Section IV) then directly reflects the module utilization. Second, we use a purely integrational controller. With a mean value of utilization assigned to the controller demand, the controller registers and cumulates all imbalances between string modules utilization. These cumulated imbalances feed the optimization routine, which consequently selects states suppressing the utilization imbalances back to zero.

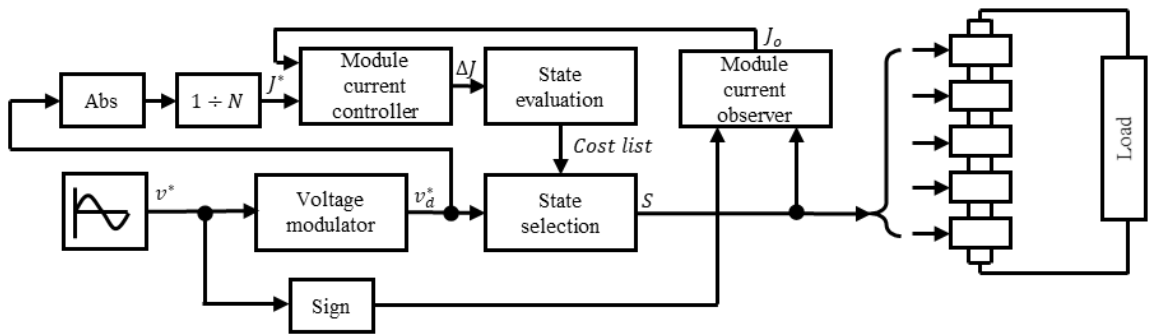

**Figure 5.** Block diagram of the control algorithm adapted for sensorless operation and open-loop control.

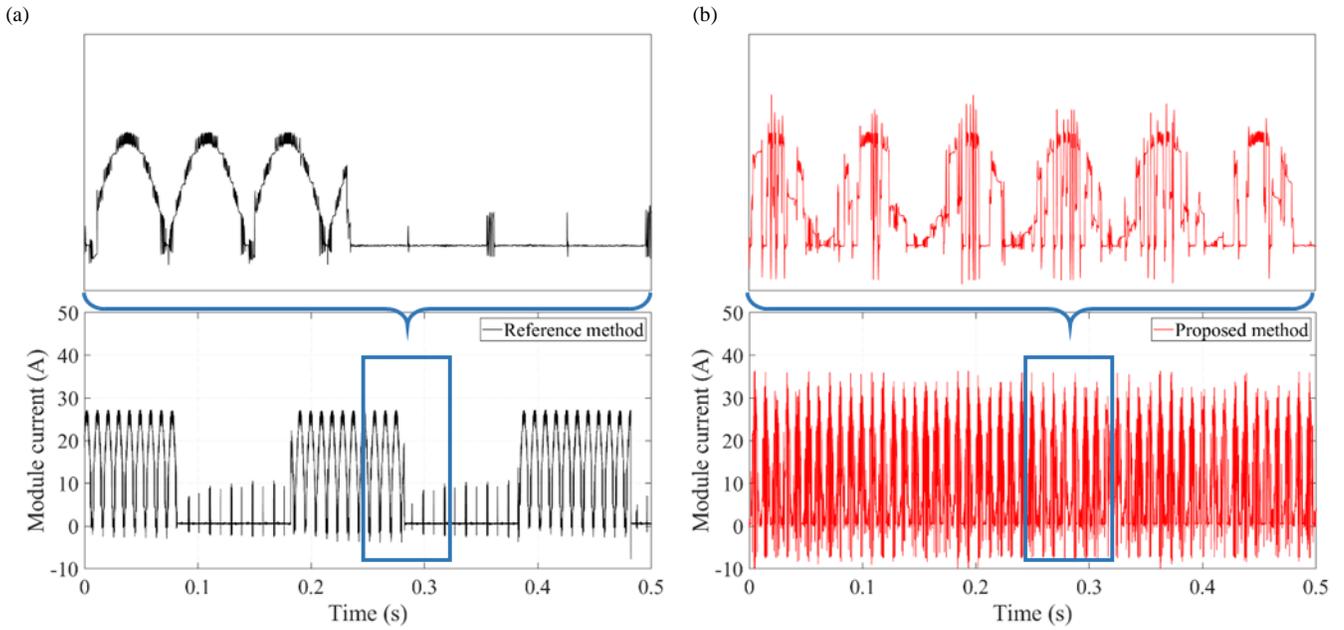

**Figure 7.** Module current waveform of (a) the reference method (with characteristic long-lasting switching patterns) and (b) the proposed method (with distributed switching).

An alternative embodiment of the sensorless operation may further increase the quality by fusing knowledge of the load power factors to accordingly shift the phase of the signum signal with respect to demanded voltage. Additionally, the signum may be substituted by the reference voltage to consider the sinusoidal character of the load. Nevertheless, precise sensorless operation is not in the scope of this paper; instead our work focusses on increased bandwidth of the scheduler sufficiently demonstrated by structure depicted in Figure 5 and its exploitation for reduced battery stress.

Additionally, we modify the control algorithm to delay the feedback loop. This modification allows us to simulate the delayed feedback according to the method proposed in [72].

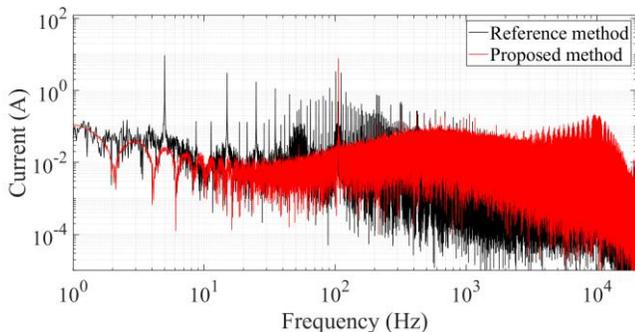

**Figure 8.** Comparison between module current spectrum of the reference and the proposed method.

The delay of the feedback is set to 100 ms, which roughly corresponds to the update rate proposed by Specht et al.

### B. Results

Figure 6 displays the experimental phase load characterized in Table I. Figure 7 presents the module current comparison between the proposed method with proposed high bandwidth and the reference method with 100 ms slacking feedback. We chose to evaluate results from module at the phase terminals, but comparable results were observed throughtout the phase string.

The delayed feedback of the reference method is noticable in the module current as load patterns (see Fig. 7a). The period length of the patterns correspond to the feedback period of 100 ms. The patterns in the module load are characterized by frequency, ranging from low up high frequency, and in extreme cases also by no load (bypassing). In case of discharging, the high frequency (or no load) is typical for lowly charged modules, when the optimization routine tries to mitigate the phase load to prevent further discharging. On the contrary, highly charged modules are preferably exposed to the phase load and experience load of lower frequency, where in extreme cases, the load of the module exactly equals the phase load. Naturally, each pattern contains a distinct number of switching events and contributes to an unequal distribution of switching in time. The switching patterns are specially harmful if their duration is longer or close to the period of the ac phase load.

Quite on the opposite, the proposed control method allows full utilization of the switching rate and evenly distributes the switching events in time. The distribution of the switching prevents extensive utilization of individual modules and naturally reduces the low frequency content (see Fig. 7b). Yet, the load of modules controlled by our method may also exhibit minor switching patterns caused by propagation delay. As long as the period length of the pattern stays negligible compared to the phase frequency of 50 Hz, the control method efficiently contributes to suppression of low-frequency content by distributed switching.

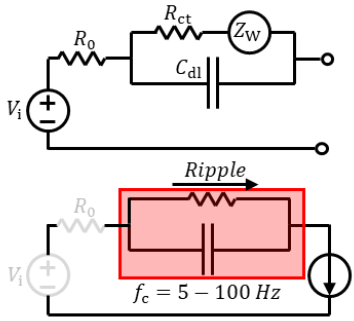

**Figure 9.** Simplified battery model: a) Randles equivalent circuit including electrochemical processes, b) first-order filter simulating behavior of double layer capacitance at higher frequencies, subsuming also the electrolyte capacitance there for simplicity.

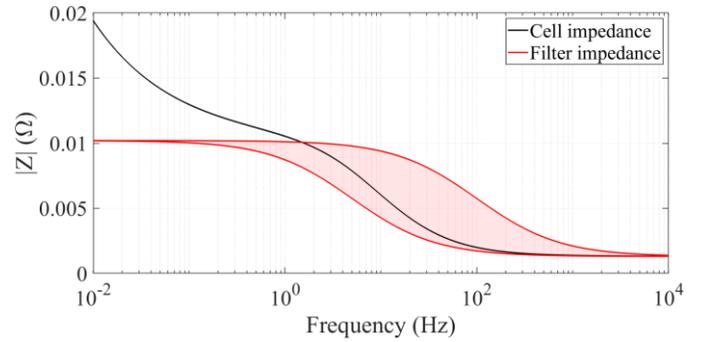

**Figure 10.** Impedance magnitudes of full Randels model in comparison with reduced first-order degradation model.

The impact of the proposed control method on the module current is more obvious in the frequency domain (see Fig. 8). The spectrum of the reference method's module current exhibits dominant peaks at twice the phase load frequency ($f = 2 \times 50 = 100$ Hz). Futher peaks can be observer at 5 Hz, which repeats at 10 Hz, 15 Hz, etc. These peaks follow from the update period of 100 ms and the patterns in the module load. It displays relatively low content at frequencies above 100 Hz, which rises again around the switching frequency.

In contrast to the prior art, the frequency content of our proposed method is practically negligible for low frequencies, just starts at 100 Hz, and features increased content up to a fraction of the switching frequency. This content is a result of appropriate switching distribution and prevention of pattern formation.

Analyzing the root mean square (RMS) values of both signals yields lower RMS to average ratio in proposed method by 3%. We attribute the lower RMS value to better utilization of the dc-link capacitors in our modules. Although our control approach does not directly aim to enhance filtering in the dc-link capacitors, the results indicate improved qualities of our approach. Still, the main contribution arguably lies in improved battery treatment.

### C. Battery degradation metric

The frequency impact of ripple currents on degradation can be supported by the frequency behavior of cells measured with electrochemical impedance spectroscopy (EIS). EIS is a technique used to identify interfacial behavior in electrochemical systems [74], where the electrical impedance is a result of different electrochemical processes and their individual contribution differs with frequency. Typically, the impedance features frequency regions characterized by dominance of individual effects, such as diffusion or dielectric capacitance (see Fig. 10). The characteristic behavior is usually approximated with existing equivalent electrical circuits.

The Randles equivalent circuit (see Fig. 9a) is derived from processes at the electrochemical interface and accordingly fits the results of the EIS. The resistance $R_0$ models the electrolyte resistance, resistance $R_{ct}$ models the charge transfer voltage drop over the electrode-electrolyte interface, and the Warburg impedance $Z_W$ models the reaction-limiting diffusion process-es. Furthermore, the components simulating the Faradaic processes are paralleled by a capacitor $C_{dl}$, which represents effect of charge building up in the electrolyte at electrode surface, also referred as double-layer capacitance [75]. For simplicity, the electrolyte capacitance, which strictly is in series with the double-layer capacitance and shunts the voltage source,

With increasing frequency, the double layer capacitance bypasses the slow diffusion-limited Faradaic processes and the impedance of the battery cell decreases inversely to the frequency. The high-pass shunting behavior is noticeable in EIS traces (see Fig. 10). Accordingly, the shunting dielectric capacitance, which avoids electrochemical reactions and their associated ageing mechanisms, is considered to explain the negligible degradation of higher frequencies.

For further analysis, we designed a small-signal approximation of the electrode interface to quantify the current components shunted by the dielectric capacitance vs. the faradaic share. The estimation of the shunting effect furthermore allows quantification of the ageing potential of various module loads and associated control methods (see Fig. 9b). We recorded module battery load currents in the experimental setup and analyzed it with this first-order approximation.

Based on measurements of $LiFePO_4$ batteries from the literature [76], the effect of the double-layer capacitance and corresponding drop in battery cell impedance is dominant in the range between 5 Hz and 1000 Hz, where the cell impedance significantly drops (see Fig 10). Similarly, experiments in [60] observe significant change in cell deterioration on the frequency range between 10 Hz and 100 Hz. We setup the cut-off frequency of the filter to range from 5 Hz to 100 Hz. Considered impedance of the high pass filter is also depicted in Fig. 10. The ageing potential corresponds to the comparison of the voltage ripple across the filter, which corresponds to the ripple of the faradaic current, and also reflects the over-potential on the electrodes, which drives the chemical processes and increases likelihood of side reactions. Both of these factors are direct cause of ageing processes and their high ripple indicates accelerated ageing of battery cells.

Figure 11 displays the voltage/current ripple (characterized by the ratio between root mean square value and the average value of the voltage/current ripple) for various values of modulation index. The error lines in the figure mark the spread between different values of the cut-off frequency, where the upper boundary corresponds to the value of ripple in a filter with cut-off frequency $f_c = 100$ Hz, the lower boundary to the $f_c = 5$ Hz, and to the $f_c = 50$ Hz.

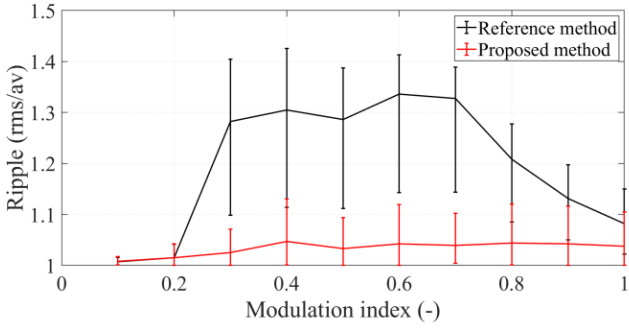

**Figure 11.** Voltage ripple ratio in the high pass filter for whole range of modulation indexes

The comparison of results of the analysis between the proposed and the reference method reveals significant suppression of the ripple in the whole range of the modulation indexes, with improvement by up to 20–30%. Whereas the reference method suffers from its strong low-frequency load caused by long repeating switching patterns, our proposed method heavily benefits from a uniform distribution of the switching. The reference method only reaches comparable performance at low modulation indices, where the system degrades to two-level operation. Further, decreasing values of the ripple ratio in reference method are also apparent at high modulation indexes, where high utilization of modules in series state reduces competence of the controller. The linearly increasing value of the ripple in the proposed method can be attributed decreasing value of effective switching frequency per module, as

$$f_{\text{module}} = \frac{f_{\text{rate}}}{N} \cdot (1 - m).$$
(6)

## VI. CONCLUSION

We presented a novel control method for battery-based MMSPC. The method aims for battery applications and their specific need for better battery treatment. Based on previous observations that the ageing potential of battery load currents decreases with frequency, we designed a control algorithm to minimize low-frequency content of the module load spectrum. Whereas the slow control loops of state-of-the-art algorithms tend to increase the low frequency content in the module load, an observer allows us to create a fast feedback loop so that our control algorithm can regulate the module currents. In addition, a simple modification of our control algorithm allows for sensorless operation.

We evaluated our novel control technique experimentally and compared it to the state of the art. Our experiment demonstrates the importance of the full exploitation of the switching frequency. A comparison of the load between a reference method from the literature and our proposed method exhibits less content in the low frequency region. The measured battery load is further evaluated in a simple battery ageing model, which discovers improved treatment of the battery cells in the system controlled by our scheduling approach.